%% file: main.tex
\documentclass[sigconf]{acmart}
\usepackage{natbib}
\usepackage{float}
\usepackage{graphicx}
\usepackage{epstopdf}
\usepackage{subfigure}
\usepackage{amsthm}
\usepackage{enumerate}
\usepackage{multirow}
\usepackage[ruled,linesnumbered]{algorithm2e}  
\usepackage{algorithmic}
\usepackage{bbm}
\usepackage{bm}
\graphicspath{{./fig/}}

\copyrightyear{2023} 
\acmYear{2023} 
\setcopyright{acmlicensed}\acmConference[SIGIR-AP '23]{Annual International ACM SIGIR Conference on Research and Development in Information Retrieval in the Asia Pacific Region}{November 26--28, 2023}{Beijing, China}
\acmBooktitle{Annual International ACM SIGIR Conference on Research and Development in Information Retrieval in the Asia Pacific Region (SIGIR-AP '23), November 26--28, 2023, Beijing, China}
\acmPrice{15.00}
\acmDOI{10.1145/3624918.3625340}
\acmISBN{979-8-4007-0408-6/23/11}

\begin{document}
\title{Unbiased Top-$k$ Learning to Rank with\\ Causal Likelihood Decomposition}


\author{Haiyuan Zhao}
\affiliation{
\institution{School of Information\\Renmin University of China}
    \city{Beijing}
    \country{China}
}
\email{haiyuanzhao@ruc.edu.cn}

\author{Jun Xu}
\authornote{Jun Xu is the corresponding author. Work partially done at Engineering Research Center
of Next-Generation Intelligent Search and Recommendation, Ministry of Education.}
\affiliation{%
  \institution{Gaoling School of Artificial Intelligence\\Renmin University of China}
    \city{Beijing}
    \country{China}
  }
\email{junxu@ruc.edu.cn}

\author{Xiao Zhang}
\affiliation{%
  \institution{Gaoling School of Artificial Intelligence\\Renmin University of China}
    \city{Beijing}
    \country{China}
  }
\email{zhangx89@ruc.edu.cn}

\author{Guohao Cai}
\affiliation{%
  \institution{Noah’s Ark Lab, Huawei}
  \city{Shenzhen}
  \country{China}
  }
\email{caiguohao1@huawei.com}

\author{Zhenhua Dong}
\affiliation{%
  \institution{Noah’s Ark Lab, Huawei}
  \city{Shenzhen}
  \country{China}
  }
\email{dongzhenhua@huawei.com}

\author{Ji-Rong Wen}
\affiliation{%
  \institution{Gaoling School of Artificial Intelligence\\Renmin University of China}
  \city{Beijing}
  \country{China}
  }
\email{jrwen@ruc.edu.cn}

\renewcommand{\shortauthors}{Haiyuan Zhao et al.}


\begin{abstract}

Unbiased learning to rank methods have been proposed to address biases in search ranking. These biases, known as position bias and sample selection bias, often occur simultaneously in real applications. Existing approaches either tackle these biases separately or treat them as identical, leading to incomplete elimination of both biases. This paper employs a causal graph approach to investigate the mechanisms and interplay between position bias and sample selection bias. The analysis reveals that position bias is a common confounder bias, while sample selection bias falls under the category of collider bias. These biases collectively introduce a cascading process that leads to biased clicks. Based on our analysis, we propose Causal Likelihood Decomposition (CLD), a unified method that effectively mitigates both biases in top-$k$ learning to rank. CLD removes position bias by leveraging propensity scores and then decomposes the likelihood of selection biased data into sample selection bias term and relevance term. By maximizing the overall log-likelihood function, we obtain an unbiased ranking model from the relevance term. We also extend CLD to pairwise neural ranking. Extensive experiments demonstrate that CLD and its pairwise neural extension outperform baseline methods by effectively mitigating both position bias and sample selection bias. The robustness of CLD is further validated through empirical studies considering variations in bias severity and click noise.



\end{abstract}
%



\maketitle

\input{1-introduction}

\input{3-relatedwork-v2}

\input{2-background-v3}
\input{4-method-v2}

\input{5-experiment}

\input{6-result}


\section{Conclusions}
In this paper, we have proposed a novel and theoretical sound model for learning unbiased ranking models in top-$k$ learning to rank, referred to as CLD. In contrast to existing methods, CLD simultaneously tackles the position bias and sampling selection biases from the viewpoint of a causal graph. It decomposes the log-likelihood function of user interactions as an unbiased relevance term plus other terms that model the biases. An unbiased ranking model can be obtained by maximizing the whole log-likelihood. Extension to the pairwise neural ranking is also developed. Experimental results verified the superiority of the proposed methods over the baselines in terms of ranking accuracy and robustness.



\begin{acks}
This work was funded by the National Natural Science Foundation of China (No. 62376275), Beijing Outstanding Young Scientist Program NO. BJJWZYJH012019100020098, the Fundamental Research Funds for the Central Universities, and the Research Funds of Renmin University of China (23XNKJ13). This work was also supported by the Intelligent Social Governance Interdisciplinary Platform, Major Innovation \& Planning Interdisciplinary Platform for the "Double-First Class" Initiative, Renmin University of China. The work was partially done at Beijing Key Laboratory of Big Data Management and Analysis Methods.
\end{acks}

\balance
\bibliographystyle{ACM-Reference-Format}
\bibliography{ref}


\end{document}


\title{Supplementary Materials for "Unbiased Top-$k$ Learning to Rank with Causal Likelihood Decomposition"}





\maketitle

\input{workplace/7-Appendix}

\balance

%% file: 1-introduction.tex
\section{Introduction}


In order to efficiently make use of user interaction data in learning of ranking models, studies on alleviating biases in user interaction data have been conducted, called Unbiased Learning to Rank (ULTR)~\citep{Ai2018Unbiased, Hu2019Unbiased, Ai2021survey} or Counterfactual Learning to Rank (CLTR)~\citep{Agarwal2019framework, Jagerman2019To,Oosterhuis2021Robust}. Previously, studies focused on position bias and usually assumed that users can examine the whole ranking list so that every relevant document is guaranteed to be examined~\citep{Joachims2017Unbiased,Carterette2018Offline,Craswell2008Experimental,Zhuang2021Cross}. Due to the limitation of the device sizes, however, search engines usually only display at most $k$ relevant documents to the user-issued query, on the basis of existing ranking models. It leads to the problem of unbiased top-$k$ learning to rank\citep{Niu2012Top}.If the ranking models are trained on the user interactions with these top-$k$ displayed documents, the sample selection bias will occur~\citep{heckman1976common, heckman1979sample}. Moreover, the user interaction data gathered from the top-$k$ ranking positions still suffers from the effect of position bias, making unbiased top-$k$ learning to rank more challenging. 

Recently,~\citet{Oosterhuis2020Policy} developed policy-aware propensity scoring to eliminate sample selection bias. They proved that the policy-aware estimator is unbiased if every relevant item has a non-zero probability to appear in the top-$k$ ranking. However, they treated position bias and sample selection bias as identical biases and applied a uniform policy-aware propensity score to reweight them. Moreover, their approach relies on knowledge of multiple logging policies and requires these policies to produce sufficiently different ranking orders, which is an expensive external intervention. In a similar vein, inspired by Heckman's two-stage method~\citep{heckman1979sample, heckman1976common}, \citet{Ovaisi2020Correcting} proposed Heckman$^\mathrm{rank}$ for top-$k$ learning to rank. To jointly correct both sample selection bias and position bias, they introduced RankAgg, which combines the results from Heckman$^\mathrm{rank}$ and IPW (Inverse Propensity Weighting). However, as position bias and sample selection bias were mitigated separately without considering their associations, the approach may still produce biased ranking outcomes. Therefore, effectively mitigating both position bias and sample selection bias simultaneously remains a challenging problem.

In this study, we analyze the biases present in user interactions using a causal graph framework. By decomposing the likelihood of top-$k$ ranking, we demonstrate that position bias is generated by the influence of examining items in different positions and represents a typical confounder bias. Conversely, sample selection bias can be understood as a collider bias. Furthermore, we identify an association between position bias and sample selection bias in the context of top-$k$ ranking. The probability of a click on a document is influenced by both its displayed position and the search engine's selection process.
These findings highlight that addressing each bias separately or treating them as identical cannot effectively produce unbiased ranking outcomes. Consequently, we are motivated to develop a novel approach capable of simultaneously mitigating both biases.

Based on our analysis, we propose a unified approach, called Causal Likelihood Decomposition (CLD), to simultaneously mitigate position bias and sample selection bias. CLD follows a cascade process to address both biases effectively.
First, CLD directly applies examination propensity scores to reweight observed user interactions, reducing the impact of position bias. However, these reweighted interactions still suffer from sample selection bias.
To further tackle sample selection bias, CLD decomposes the log-likelihood function into distinct components. One component represents an unbiased term solely based on user-perceived relevance, while the other components capture sample selection bias. This decomposition allows for detachment of relevance signals from the observed user interaction data.
Theoretical analysis demonstrates that by maximizing the entire log-likelihood function, an unbiased relevance ranking model can be obtained from the unbiased term.
Furthermore, we present an extension of CLD that incorporates neural networks as ranking and selection models. The parameters of these models are learned using pairwise losses, enhancing their effectiveness in dealing with bias mitigation.

CLD offers several advantages: 
theoretical soundness, elegant extension to pairwise neural ranking, and high accuracy in unbiased top-$k$ learning to rank.
The major contributions of this work are:
\begin{enumerate}[(1)]
\item A theoretical analysis towards position bias and sample selection bias from the viewpoint of statistical causal inference;
\item A unified and theoretical sound approach to mitigating both position bias and sample selection bias in top-$k$ ranking. The method is derived under likelihood maximization and can be applied to both pointwise and pairwise training;
\item Extensive experiments on two publicly available datasets demonstrated the effectiveness of the proposed approaches over baselines for the task of unbiased top-$k$ learning to rank. The empirical analysis also showed the robustness of the approaches in terms of the variation of bias severity and the click noise. 
\end{enumerate}

%% file: 3-relatedwork-v2.tex
\section{Related work}
\subsection{Unbiased learning to rank}
Recently, there has been a trend towards utilizing the user interaction data (e.g., the click log) as the substitute for the expert annotated relevance labels to train the ranking models in web search~\citep{Yuan2020Unbiased, Hu2019Unbiased, Wang2016Learning}. In contrast to the expert annotated labels, the user interaction data is massive, cheap, and most importantly, user-centric~\citep{Agarwal2019framework}. However, the behaviors of the users could probably be affected by some unexpected factors~\citep{Joachims2005Accurately}, including the display ranking position~\citep{Joachims2007Evaluating, Joachims2017Unbiased, Ai2018Unbiased, Agarwal2019Estimating, Agarwal2019framework, Fang2019Intervention, dong2020counterfactual}, the search engine's selection~\citep{Ovaisi2020Correcting, Oosterhuis2020Policy, Ovaisi2021Propensity}, and others\citep{schnabel2016recommendations, wang2019doubly, Saito2020Asymmetric,Agarwal2019Addressing, Vardasbi2020When, dong2020counterfactual, Wei2021Model}.
These factors, along with users' true perceived relevance, impact 
the observational user interaction data gathered from search engines. Although relevance signal is contained inside, the interaction data cannot be directly used to train the ranking models unless those aforementioned factors are eliminated. Otherwise, a biased ranking model would be learned and hurt the user experience~\citep{chen2020survey}.

To mitigate biases, unbiased learning to rank has attracted a lot of research efforts recently. Most approaches focus on addressing a single bias. 
For example, inverse propensity weighting (IPW) has been widely discussed in many studies~\citep{Joachims2017Unbiased,Agarwal2019framework} for addressing the position bias. It estimates the causal effect of examination and extracts them from the click signal directly. The top-$k$ cut-off of search engines leads to sample selection bias.  Inspired by the famous Heckman’s two-stage method~\citep{heckman1979sample, heckman1976common}, \citet{Ovaisi2020Correcting} proposed $Heckman^{rank}$ for top-$k$ learning to rank. 
Furthermore,~\citet{Oosterhuis2020Policy} developed policy-aware propensity scoring to eliminate sample selection bias, by assuming that policy-aware estimator knows multiple differ enough logging policies.

In the real world, various biases could occur simultaneously~\citep{chen2020survey}. To address the trust bias and position bias simultaneously,~\citet{Agarwal2019Addressing} proposed a Bayes-IPS estimator. Affine-IPS~\citep{Vardasbi2020When} improved Bayes-IPS and achieved better performance.~\citet{Ovaisi2020Correcting} proposed RankAgg who ensembles the ranking results by the model for correcting position bias and the model for correcting sample selection bias.~\citet{Ovaisi2021Propensity} proposed PIJD which does not require the exact propensity scores and can mitigate both position bias and sample selection bias. More recently,~\citet{Oosterhuis2021Unifying} introduced an intervention-aware estimator for integrating counterfactual and online learning to rank, which can mitigate position bias, sample selection bias, and trust bias simultaneously. \citet{Chen2021AutoDebias} proposed AutoDebias that leverages the uniform data to learn the optimal debiasing strategy for various biases.

\subsection{Causal inference in information retrieval} 
The generation of users' implicit feedback in real search engines is affected by many biased factors. To make this feedback usable, causal inference has been introduced to analyze the generation procedure of users' implicit feedback and mitigate bias inside. For instance,~\citet{Zheng2021Disentangling} analyzed the casual structure of popularity bias and proposed DICE to disentangle the user interest from click.~\citet{Zhang2021Causal} further analyzed the causal structure of item popularity and leveraged them to enhance the performance of recommendation.~\citet{Wang2020Causal} utilized the causal inference to handle the unobserved confounders in the recommendation. ~\citet{Wang2021Deconfounded} proposed DecRs to dynamically regulate backdoor adjustment according to user status, thus eliminating the effect of confounders. However, few works conduct a systematical analysis for the bias in ranking from the viewpoint of causal inference.

%% file: 2-background-v3.tex
\section{\mbox{Analyzing the Biases in top-k ranking}}
\subsection{Problem Formulation}
The problem of unbiased top-$k$ learning to rank can be described as follows. Given a user query $q$ and $K$ retrieved documents, each query-document pair $(q,d)$ is described by a feature vector $\mathbf{x}=\phi(q,d) \in\mathbb{R}^n$. The relevance of $(q,d)$ can be represented by an unobserved variable $R$, which could be binary, ordinal, or real. The retrieved documents are ranked by a logging policy (an existing ranking model) $\pi_0: \mathbb{R}^n\mapsto \{1, 2, \cdots, K\}$ according to their features, where each document will be ranked at some position $P\in \{1, 2, \cdots, K\}$ by $\pi_0$. In the real world, only the top $k \leq K$ documents can be presented to users due to some limitations (e.g., screen sizes). Let's use $S\in \{0, 1\}$ to denote whether a document is selected and presented to the user, i.e., $S=1$ if selected otherwise not. Further, let's use $E\in \{0, 1\}$ to denote that the user has examined the presented document, and $C\in \{0,1\}$ to denote whether a user clicks the document, which is a random variable obeying Bernoulli distribution\citep{Joachims2017Unbiased, Agarwal2019framework, Vardasbi2020When}. 

\begin{equation}
    \begin{split}
        \mathcal{L}_{\mathrm{biased}} &= \sum_{i =1}^{N_s} \log\left(\mathrm{Pr}(c_i|\mathbf{x}_i)\right),
    \end{split}
    \label{gjk}
\end{equation}

\begin{equation}
    \begin{split}
        \mathcal{L}_{\mathrm{de.}\;\mathrm{both}\;\mathrm{biases}} &= \sum_{i=1\land s_{i}=1}^{N_s}\log\left(\mathrm{Pr} \left(\mathbb{E}\left[\frac{c_i}{\rho_i}\right] \bigg|\mathbf{x}_{i} \right)\right)\\
        &+\sum_{i=1\land s_{i}=1}^{N_s}\log\left(\mathrm{Pr} \left(s_i \bigg|~\mathbb{E}\left[\frac{c_i}{\rho_i}\right],\mathbf{x}_{i} \right)\right)\\
        &+\sum_{i=1\land s_{i}=0}^{N_u}\log\left(\mathrm{Pr}(s_{i}|\mathbf{x}_{i})\right),
    \end{split}
    \label{eq: nobothbiases_likelihood}
\end{equation}

The user interactions with a search engine can be recorded as click log $\mathcal{D}=\{(\mathbf{x}_i, c_{i}, k_{i}, s_{i})\}_{i=1}^{N}$, where $\mathbf{x}_i, c_{i}, k_{i}, s_{i}$ respectively denote the $i$-th query-document pair's feature vector, whether the document being clicked, the rank position, and whether being selected. 
Ideally, we hope an unbiased ranking model could be estimated by maximizing the log-likelihood shown below:
\begin{equation}
    \begin{split}
        \mathcal{L}_{\mathrm{unbiased}} &= \sum_{i =1}^N  \log\left(\mathrm{Pr}(r_i|\mathbf{x}_i)\right),
    \end{split}
    \label{eq: unbias_lilelihood_point}
\end{equation}
where the $r_i$ is the unobserved relevance for query-document pairs encoded by $\mathbf{x}_i$. Equation~(\ref{eq: unbias_lilelihood_point}) cannot be maximized because $r_i$ cannot be observed directly.

On the other hand, we observed that the click log consists of two parts: $\mathcal{D} = \mathcal{D}_s \bigcup \mathcal{D}_u$, where  $\mathcal{D}_s=\{(\mathbf{x}_i, c_i, k_i, s_i = 1)\}_{i=1}^{N_s}$ are the interactions for the $N_s$ selected $(q,d)$ pairs, and $\mathcal{D}_u=\{(\mathbf{x}_i, c_i = 0, k_i, s_i = 0)\}_{i=1}^{N_u}$ are those for the $N_u$ not selected pairs. A naive log-likelihood can be written as:
\begin{equation}
    \begin{split}
        \mathcal{L}_{\mathrm{naive}} &= \sum_{i=1}^{N_s} \log\left(\mathrm{Pr}(c_{i}|s_{i}=1,\mathbf{x}_{i})\mathrm{Pr}(s_{i}=1|\mathbf{x}_{i})\right) \\ &+\sum_{i=1}^{N_u}\log\left(\mathrm{Pr}(s_{i}=0|\mathbf{x}_{i})\right).
    \end{split}
    \label{eq: naive_likelihood_point}
\end{equation}
Though it can be optimized, the naive log-likelihood suffers from both position bias from $c_{i}$ and the sample selection bias from $s_{i}$. There exists a large gap between the naive Equation~\eqref{eq: naive_likelihood_point} and the ideal unbiased objective Equation~\eqref{eq: unbias_lilelihood_point}. Table~\ref{tab:notation} lists the major notations in the paper.

\begin{figure}[t]
    \includegraphics[width=0.36\textwidth]{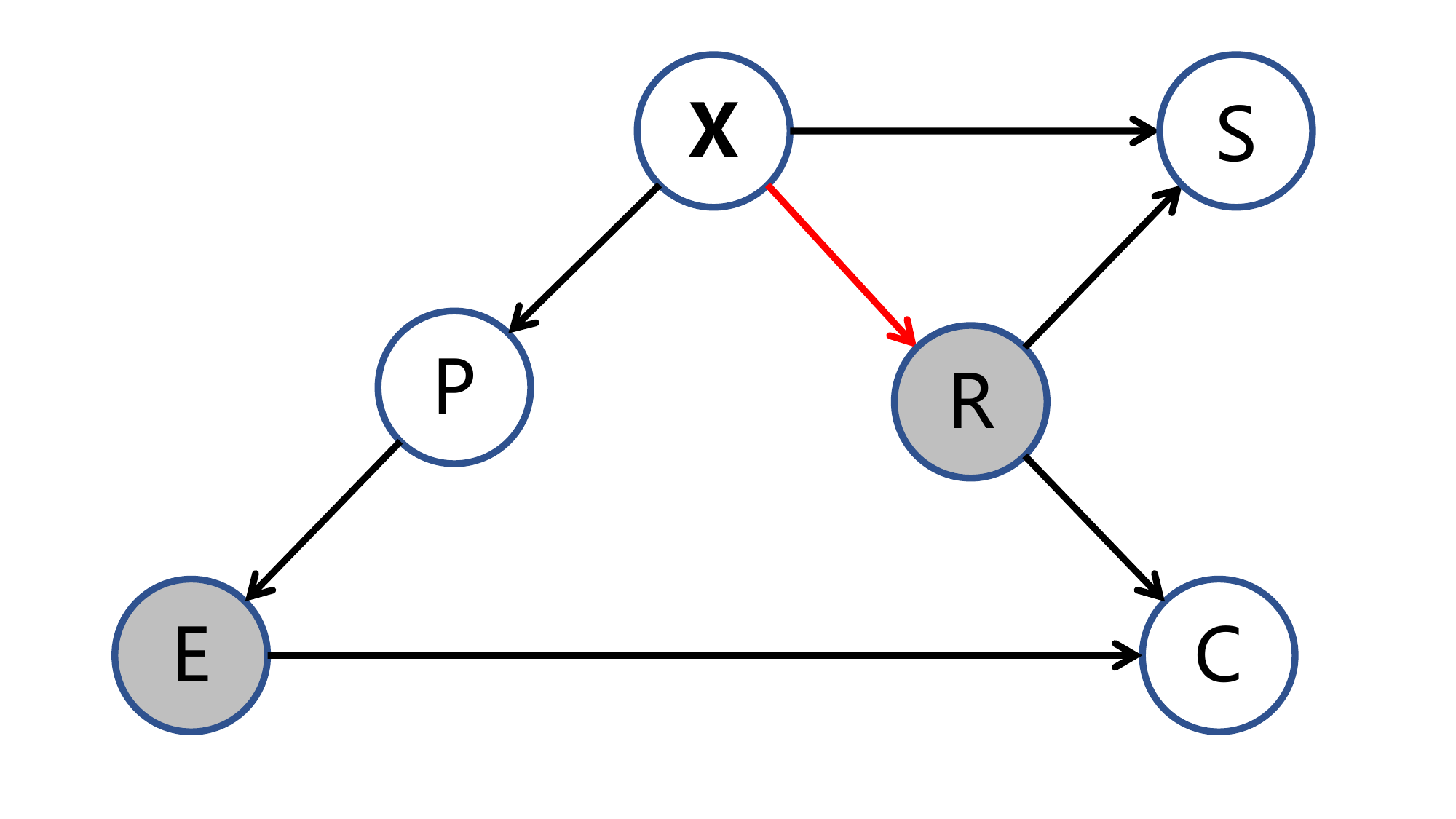}
    \caption{The casual graphs of observed log data. Each node corresponds a casual variable and the gray node means that the variable is unobserved. The red arrow ($\mathbf{x} \rightarrow R$) denote the effect that an unbiased ranking model needs to estimate.}
    \label{fig:biascausalgraph}
\end{figure}

\begin{table}[t]
    \caption{Notations and explanations.}
    \label{tab:notation}
        \begin{tabular}{cl}
        \toprule
            Notation& Description\\
        \midrule
            $(q,d)$ & a query-document pair\\
            $\mathbf{x}=\phi(q,d)$ & feature vector in $\mathbb{R}^n$ corresponding to a $(q,d)$ pair\\
            $R,\;r_i$ & true relevance of a $(q,d)$ pair (unobserved)\\
            $E$ & user's examination on a document (unobserved)\\
            $C,\;c_i$ & click on a document (can be observed)\\
            $P,\;k_i$ & position of a document being displayed (observed)\\
            $S,\;s_i$ & whether a document being selected (observed)\\
            $\mathcal{D}=\mathcal{D}_s \bigcup \mathcal{D}_u$ & click log for selected and not selected $(q,d)$ pairs\\
            $\rho_i$ & the propensity score of the $i$-th $(q,d)$ pair\\
            $\pi_0$ & logging policy (an existing ranking model)\\
        \bottomrule
        \end{tabular}
\end{table}

\subsection{Causal view of the biases in top-$k$ ranking }
\label{sec: causalview_position}
Next, we will illustrate why ranking models are biased if they are trained with user clicks directly, and reveal the distinctions and association between the position bias and sample selection bias. In order to accomplish this, we introduce a causal graph representing the observed log data, depicted in Figure~\ref{fig:biascausalgraph}. This causal graph comprises six causal variables denoted as $\{\mathbf{x},P,R,S,E,C\}$ which have been defined in Table~\ref{tab:notation}. The graph's edges describe causal relations between variables:
\begin{itemize}
    \item $\mathbf{x} \rightarrow R$: this edges represents the causal relations of the query- document pair's feature $\mathbf{x}$ and their corresponding relevance score $R$, which is the effect that an unbiased ranking model needs to estimate.
    \item $\mathbf{x} \rightarrow P$: the displayed position $P$ is determined by a logging policy $\pi_0(\mathbf{x})$ which takes the feature $\mathbf{x}$ as the input.
    \item $P \rightarrow E$: the chance of an item examined by users is determined by its displayed position $P$, note that the variable $E$ is unobserved.
    \item $(\mathbf{x}, R) \rightarrow S$: the rank position used to select is determined by the predicted score of logging policy $\pi_0$, and $\pi_0$ is trained by $\mathbf{x}$ and $R$\footnote{In real search practices, a small number of the human label can be utilized to train $\pi_0$.}, thus the selection $S$ is indirectly determined by $\mathbf{x}$ and $R$. The intermediate factors are omitted for simplifying the illustration.
    \item $(E,R) \rightarrow C$: the click $C$ on a query-document pair will be impacted by both the examination $E$ and the perceived relevance $R$ of a user, which is followed by the examination hypothesis~\citep{Richardson2007Predicting}
\end{itemize}

Drawing upon causal inference principles~\citep{pearl2009causality, Pearl2016Causal}, we identify two types of biases present in our causal graph, aligning with the previously mentioned position bias and sample selection bias:
\begin{itemize}
    \item $E \leftarrow \mathbf{x} \rightarrow C$ (\emph{confounder bias}): The feature vector $\mathbf{x}$ of query-document pair is the confounder of examination $E$ and click $C$, which leads to a spurious correlation between $E$ and $C$. Since $E$ is determined by rank position, the confounder bias is presented as position bias in top-$k$ ranking.
    \item $\mathbf{x} \rightarrow S \leftarrow R$ (\emph{collider bias}): $S$ is the collider~\citep{pearl2009causality} between $\mathbf{x}$ and $R$. The backdoor path between $\mathbf{x}$ and $R$ will be opened when conditioned on $S$. The collider bias manifests as sample selection bias in top-$k$ ranking~\citep{cole2010illustrating}.
\end{itemize}
It is worth noting that some studies also discuss the trust bias\citep{Agarwal2019Addressing, Vardasbi2020When}, which introduces an additional causal relation $P \rightarrow C$. However,  trust bias can also be considered a form of confounder bias, sharing a similar mechanism with position bias. Since our primary focus is on addressing both confounder bias and collider bias, we have chosen to omit the trust bias in this study. From the analysis presented above, it becomes evident that position bias and sample selection bias in top-$k$ ranking are distinct biases that require separate approaches for mitigation.

Meanwhile, the position bias and sample selection bias also have association according to Figure~\ref{fig:biascausalgraph}. Note that the existance of causal relations $R \rightarrow C$ bridge the association of the path of position bias ($E \leftarrow \mathbf{x} \rightarrow C$) and sample selection bias ($\mathbf{x} \rightarrow S \leftarrow R$). This implies that the relevance R initially experiences sample selection bias, which subsequently amplifies the spurious correlation introduced by position bias through $R \rightarrow C$. As a result, $C$ becomes influenced by both biases simultaneously. To provide further clarity, the selected click probability $\Pr (C=1\mid S=1, \mathrm{x})$ (a component of Equation~(\ref{eq: naive_likelihood_point}), which is abbreviated as $p_C^{S=1}$ in the following equation) can be decomposed as follows:
\begin{align}
        p_C^{S=1}&=\sum_{E}\sum_{R}\mathrm{Pr}(C=1|E,R) \mathrm{Pr}(R|S=1,\mathbf{x}) \sum_{i=1}^K \mathrm{Pr}(E|P=i)\mathrm{Pr}(P=i|\mathbf{x}) \nonumber \\
        &=\sum_{E}\sum_{R}\mathrm{Pr}(C=1|E,R) \mathrm{Pr}(R|S=1,\mathbf{x}) \mathrm{Pr}(E|P=k)\nonumber\\
        &=\sum_{R}\mathrm{Pr}(C=1|E=1,R) \mathrm{Pr}(R|S=1,\mathbf{x}) \mathrm{Pr}(E=1|P=k)\nonumber \\
        &=\underbrace{\mathrm{Pr}(E=1|P=k)}_{\text{Position Bias}} \underbrace{\mathbb{E}_{R\sim\mathrm{Pr}(R|S=1,\mathbf{x})}[\sigma(R)]}_{\text{Selection biased relevance}} ,
    \label{eq: likelihood_decomp}
\end{align}

where $k=\pi_0(\mathbf{x})$. 
The first line is an expansion based on Figure~\ref{fig:biascausalgraph}. The second line is based on the assumption that $\pi_0$ is a deterministic ranking policy, meaning that one document in a query only has one corresponding rank position. The third line is according to the examination hypothesis~\citep{Richardson2007Predicting}. In the last line, we treat the click probability $\mathrm{Pr}(C=1|E=1,R)$ as a function of relevance $R$, denote as $\sigma(R)$. Here $\sigma(\cdot)$ is a monotonically increasing function for mapping $R$ to the interval 0 to 1 (e.g. Sigmoid function). This mapping serves as an indicator for relevance $R$. The decomposition equation (Equation~\ref{eq: likelihood_decomp}) demonstrates that user-perceived relevance is initially impacted by sample selection bias and subsequently influenced by position bias, forming a cascade process. To obtain an unbiased relevance signal, it is essential to first eliminate the position bias in the click signal and then mitigate the sample selection bias.

%% file: 4-method-v2.tex
\section{Unified bias mitigation in top-$k$ ranking}
\label{sec: point method}
Based on the analysis in the above section, this section presents a model called Causal Likelihood Decomposition (CLD) which simultaneously mitigates the position bias and the sample selection bias in top-$k$ learning to rank. 

\subsection{Formulation of the log-likelihood}
\label{sec: formulation}
The analysis of Equation~(\ref{eq: likelihood_decomp}) in Section~\ref{sec: causalview_position} indicates that the click signals can be transformed to selection biased relevance with the help of examination propensity $\mathrm{Pr}(E=1|P=k)$: 
\[
\mathbb{E}_{R\sim\mathrm{Pr}(R|S=1,\mathbf{x})}[\sigma(R)] = \frac{\Pr (C=1\mid S=1, \mathrm{x})}{\mathrm{Pr}(E=1|P=k)} = \mathbb{E}\left[\frac{C}{\mathrm{Pr}(E=1|P=k)}\right]
\]

Therefore, the click in Equation~\eqref{eq: naive_likelihood_point} can be replaced with the expectation of propensity re-weighted click, achieving a log-likelihood with position bias be detached:
\begin{equation}
        \mathcal{L}_{\mathrm{de.}\;\mathrm{posi.}} = \!\!\!\!\sum_{i=1\land s_i=1}^{N_s}\!\!\!\!\log\left(\mathrm{Pr}\left(\mathbb{E}\left[\frac{c_i}{\rho_i}\right],s_{i} \bigg|\mathbf{x}_{i} \right)\right)
        +\!\!\!\!\sum_{i=1\land s_i=0}^{N_u}\!\!\!\!\log\left(\mathrm{Pr}(s_{i}|\mathbf{x}_{i})\right),
    \label{eq: nopositionbias_likelihood}
\end{equation}
where $\mathbb{E}\left[\frac{c_i}{\rho_i}\right]=\mathbb{E}\left[\frac{C}{\mathrm{Pr}(E=1|P=k)}\right]$ for simplifying the notations. Note that $\rho_i$ is the examination propensity score for the $i$-th $(q,d)$ pair. A number of studies have been proposed to estimate the weights~\citep{Agarwal2019Estimating, Ai2018Unbiased, Wang2018Position}. In this paper, we treat them as known values.

Equation~\eqref{eq: nopositionbias_likelihood} still suffers sample selection bias because the first term still contains $s_i$. Therefore, the likelihood of selected biased data can be further decomposed to contain the unbiased learning target, thus detaching sample selection bias: 
\begin{equation}
    \begin{split}
        \mathcal{L}_{\mathrm{de.}\;\mathrm{both}\;\mathrm{biases}} &= \sum_{i=1\land s_{i}=1}^{N_s}\log\left(\mathrm{Pr} \left(\mathbb{E}\left[\frac{c_i}{\rho_i}\right] \bigg|\mathbf{x}_{i} \right)\right)\\
        &+\sum_{i=1\land s_{i}=1}^{N_s}\log\left(\mathrm{Pr} \left(s_i \bigg|~\mathbb{E}\left[\frac{c_i}{\rho_i}\right],\mathbf{x}_{i} \right)\right)\\
        &+\sum_{i=1\land s_{i}=0}^{N_u}\log\left(\mathrm{Pr}(s_{i}|\mathbf{x}_{i})\right),
    \end{split}
    \label{eq: nobothbiases_likelihood}
\end{equation}
Same as the naive likelihood in Equation~\eqref{eq: naive_likelihood_point}, Equation~\eqref{eq: nobothbiases_likelihood} is still the likelihood among all observed data. The difference is both position bias and sample selection bias have been detached, and the unbiased learning target has been decomposed as the first term of Equation~\eqref{eq: nobothbiases_likelihood}.

\begin{equation}
    \begin{split}
        \mathcal{L}_{\mathrm{de.}\;\mathrm{both}\;\mathrm{biases}} &= \sum_{i=1\land s_{i}=1}^{N_s}\log\left(\mathrm{Pr} \left(\mathbb{E}\left[\frac{c_i}{\rho_i}\right] \bigg|\mathbf{x}_{i} \right)\right)\\
        &+\sum_{i=1\land s_{i}=0}^{N_u}\log\left(\mathrm{Pr}(s_{i}|\mathbf{x}_{i})\right),
    \end{split}
    \label{eq: nobothbiases_likelihood}
\end{equation}



\begin{algorithm}[t]

    \SetAlgoLined
    \KwIn{iteration number $T$, click log
    $\mathcal{D} = \mathcal{D}_{s}\bigcup \mathcal{D}_u$
    }
    
    \KwOut{Model parameters $\bm{\beta}$ and $\bm{\omega}$}
    
$\bm{\rho} \leftarrow$ estimate $K$ propensity scores; 

%
$\bm{\beta},\bm{\omega} \leftarrow$ Xavier initialization\citep{glorot2010understanding};\\
    \For{$1\leq t\leq T$}
    {
        Randomly sample a batch of sessions $\mathcal{D^{'}}$ from $\mathcal{D}_s\cup\mathcal{D}_u$\;
        \For{$\left(\mathbf{x}_{i},c_i, k_i, s_{i}\right)\in\mathcal{D^{'}}$}
        {
            $\rho_i\leftarrow \bm{\rho}[k_i]$
            
            \eIf{$s_{i}=1$}
            {
                Update $\bm{\beta},\bm{\omega}$ with the gradient of Eq.~\eqref{eq: MULE_point} \;
            }
            {
                Update $\bm{\omega}$ with the gradient of Eq.~\eqref{eq: MULE_point}\;
            }
        }
    }
    
    \Return{$\bm{\beta},\bm{\omega}$}
    \caption{The training procedure of CLD}
    \label{alg: point_training}
\end{algorithm}

\subsection{Optimization}
\label{sec: Optimization}
To optimize Equation~\eqref{eq: nobothbiases_likelihood}, we followed the \emph{Type II Tobit model}~\cite{Amemiya1984Tobit}, which parameterized the likelihood in Equation~\eqref{eq: nobothbiases_likelihood} under the linear and Gaussian assumptions. Specifically, assuming that both the selection model and the ranking model are linear:
\begin{equation*}
    s_{i}=
    \begin{cases}
    0&\;\mbox{if}\; \mathbf{x}_{i}^{T}\bm{\omega}+\epsilon_{i}\leq 0\\
    1&\;\mbox{if}\;\mathbf{x}_{i}^{T}\bm{\omega}+\epsilon_{i}>0
    \end{cases},\quad
    r_{i}=
    \begin{cases}
    \mathbf{x}_{i}^{T}\bm{\beta}+\mu_{i}
    &\;\mbox{if}\; s_{i}=1\\
    \mbox{unobserved}&\;\mbox{if}\;s_{i}=0
    \end{cases},
\end{equation*}
where $s_{i}$ and $r_{i}$ are the selection indicator and the selection biased relevance of the $i$-th $(q,d)$, respectively, and both of them are calculated based on the feature vector $\mathbf{x}_{i}$. $\bm{\omega}$ and $\bm{\beta}$ are the parameters of these two linear models. $\epsilon_{i}$ and $\mu_{i}$ are the I.I.D. noises that obey Gaussian distributions and their variances are assumed to be 1. 

According to the derivations and conclusions in \citep{Amemiya1984Tobit}, the parameterized log-likelihood of Equation~\eqref{eq: nobothbiases_likelihood} becomes:
\begin{equation}
    \begin{split}
        \mathcal{L}_{\mathrm{CLD}}(\bm{\beta},\bm{\omega}) =& -\sum_{i=1\land s_{i}=1}^{N_s} \left(\mathbb{E}\left[\frac{c_i}{\rho_i}\right]-\mathbf{x}_{i}^{T}\bm{\beta} \right)^2\\
        &+\sum_{i=1\land s_{i}=1}^{N_s}\log\Phi\left(\frac{\mathbf{x}_{i}^{T}\bm{\omega}+\gamma \left(\mathbb{E}\left[\frac{c_i}{\rho_i}\right]-\mathbf{x}_{i}^{T}\bm{\beta} \right)}{(1-\gamma^{2})^{\frac{1}{2}}}\right)\\
        &+\sum_{i=1\land s_{i}=0}^{N_u}\log\left(1-\Phi(\mathbf{x}_{i}^{T}\bm{\omega})\right),
    \end{split}
    \label{eq: MULE_point}
\end{equation}
where $\Phi$ is the cumulative distribution function of a standard normal distribution, $\gamma$ is the correlation coefficient of the error terms of $\epsilon_i$ and $\mu_i$, which indicates how the selection of a $(q,d)$ pair related to its relevance. In our implementation, $\gamma$ is treated as a hyper parameter. 
Maximizing Equation~\eqref{eq: MULE_point}
achieves an unbiased estimation of $\bm{\beta}$ and $\bm{\omega}$:
\[
    (\bm{\beta^{*}},\bm{\omega^{*}}) \leftarrow \mathop{\arg\max}_{\bm{\beta},\bm{\omega}}~\mathcal{L}_{\mathrm{CLD}}(\bm{\beta},\bm{\omega}).
\]

Algorithm~\ref{alg: point_training} shows the procedure of the CLD learning algorithm for learning unbiased relevance ranking model $\bm{\beta}$ (and the selection model $\bm{\omega}$). The inputs to the algorithm are click log with  feature, selection indicator, and propensity score re-weighted click signals. After sampling a batch of data, the algorithm updates both models if this data record was selected, and only updates the selection model if it was not selected.

\subsection{Online ranking}
The outputs of the learning algorithm are the parameters of the ranking model $\bm{\beta^{*}}$ and parameters of the selection model $\bm{\omega^{*}}$. Intuitively, the selection model is used to absorb the sample selection bias while the relevance model is used to obtain the unbiased estimation of relevance. Therefore, in online ranking, a $(q,d)$ pair's ranking score is calculated as an unbiased estimation of relevance:
$$
\hat{r} = \langle \phi(q,d), \bm{\beta^{*}}\rangle.
$$

\begin{algorithm}[t]
    \SetAlgoLined
    \KwIn{iteration number $T$, click log $\mathcal{D} = \mathcal{D}_{s}\bigcup \mathcal{D}_u$}
    \KwOut{Model parameters $\bm{\beta}$, $\bm{\omega}$}
    \tcp*[h]{Create preference pairs based on $\mathcal{D}$}\;
        
    $\bm{\rho} \leftarrow$ estimate $K$ propensity scores; 
    
\mbox{$\mathcal{D}_{s}^{pair}\!\!\!\!\leftarrow\left\{\left((\mathbf{x}_{i},s_{i}),(\mathbf{x}_{j},s_{j})\right)\Big|~\mathbb{E}\left[\frac{c_i}{\bm{\rho}[k_i]}\right]>\mathbb{E}\left[\frac{c_j}{\bm{\rho}[k_j]}\right] ,s_i=1\land s_j=1 \right\}$\;}
    
    $\mathcal{D}_{u}^{pair}\!\!\!\!\leftarrow\left\{\left((\mathbf{x}_{i},s_{i}),(\mathbf{x}_{j},s_{j})\right)\Big|~s_i=0\lor s_j=0 \right\}$\;

    $\bm{\beta},\bm{\omega} \leftarrow$ Xavier initialization\citep{glorot2010understanding}\;
    \For{$1\leq t\leq T$}
    {
        Randomly sample a batch $\mathcal{D^{'}}$ from $\mathcal{D}_s^{pair}\cup\mathcal{D}_u^{pair}$\;
        \For{$\left((\mathbf{x}_{i},s_{i}),(\mathbf{x}_{j},s_{j})\right)\in\mathcal{D^{'}}$}
        {
            \eIf{$s_{i}=1\lor s_{j}=1$}
            {
                Update $\bm{\beta},\bm{\omega}$ with the gradient of Eq.~\eqref{eq: MULE_pair} \;
            }
            {
                Update $\bm{\omega}$ with the gradient of Eq.~\eqref{eq: MULE_pair}\;
            }
        }
    }
    \Return{$\bm{\beta},\bm{\omega}$}
    \caption{Pairwise training for CLD}
    
    \label{alg: pair_training}
\end{algorithm}

\section{\mbox{Extension to pairwise neural ranking}}
\label{sec: pair method}
The models learned by Algorithm~\ref{alg: point_training} are limited to be linear and learned with a pointwise objective function. Previous studies have shown that the neural ranking models learned with a pairwise objective such as BPR~\citep{rendle2012bpr} usually achieve better results. In this section, we extend the proposed pointwise and linear CLD model to pairwise neural ranking, denoted as CLD$^{\textrm{pair}}$. 

To derive the pairwise format of CLD, we first give the unbiased log-likelihood in pairwise format:
\begin{equation}
    \begin{split}
        \mathcal{L}_{\mathrm{unbiased}}^{\mathrm{pair}} &= \sum_{r_{i}>r_{j}}\log\left(\mathrm{Pr}(r_{i}>r_{j}|\mathbf{x}_{i},\mathbf{x}_{j})\right).
    \end{split}
    \label{eq: unbias_likelihood_pair}
\end{equation}
For a document $i$ and document $j$ in the ranking list of query $\bm{q}$, the pairwise unbiased likelihood is consists of the relative order of their relevance comparisons. Unfortunately, the relevance $r_{i}$ and $r_{j}$ is unknown for us. What we can observe is the click signal of each document, Based on Equation~\eqref{eq: nobothbiases_likelihood} and Equation~\eqref{eq: unbias_likelihood_pair}, the decomposed log-likelihood in pairwise can be written as
\begin{align}
        \mathcal{L}_{\mathrm{de.}\;\mathrm{both}\;\mathrm{biases}}^{\mathrm{pair}} &= \sum_{\bar{r}_{i}>\bar{r}_{j},\;s_{i}=1\land s_{j}=1}\log\left(\mathrm{Pr}(\bar{r}_{i}>\bar{r}_{j}|\mathbf{x}_{i},\mathbf{x}_{j})\right) \nonumber\\
        &+\sum_{\bar{r}_{i}>\bar{r}_{j},\;s_{i}=1\land s_{j}=1}\log\left(\mathrm{Pr}(s_{i},s_{j}|\bar{r}_{i}>\bar{r}_{j},\mathbf{x}_{i},\mathbf{x}_{j})\right) \nonumber\\
        &+\sum_{s_{i}=0\lor s_{j}=0}\log\left(\mathrm{Pr}(s_{i},s_{j}|\mathbf{x}_{i},\mathbf{x}_{j})\right), 
    \label{eq: nobothbias_likelihood_pair}
\end{align}
where $\bar{r}_{i}=\mathbb{E}\left[ c_i / \rho_i \right]$ and $\bar{r}_{j}=\mathbb{E}\left[ c_j / \rho_j \right]$.  
Since $\mathbb{E}\left[ c_i / \rho_i \right]= r_i$, the first term of Equation~\eqref{eq: nobothbias_likelihood_pair} implies an unbiased log-likelihood. Maximizing Equation~\eqref{eq: nobothbias_likelihood_pair} can obtain an unbiased ranking model.

\begin{figure}[t]
    \includegraphics[scale=0.25]{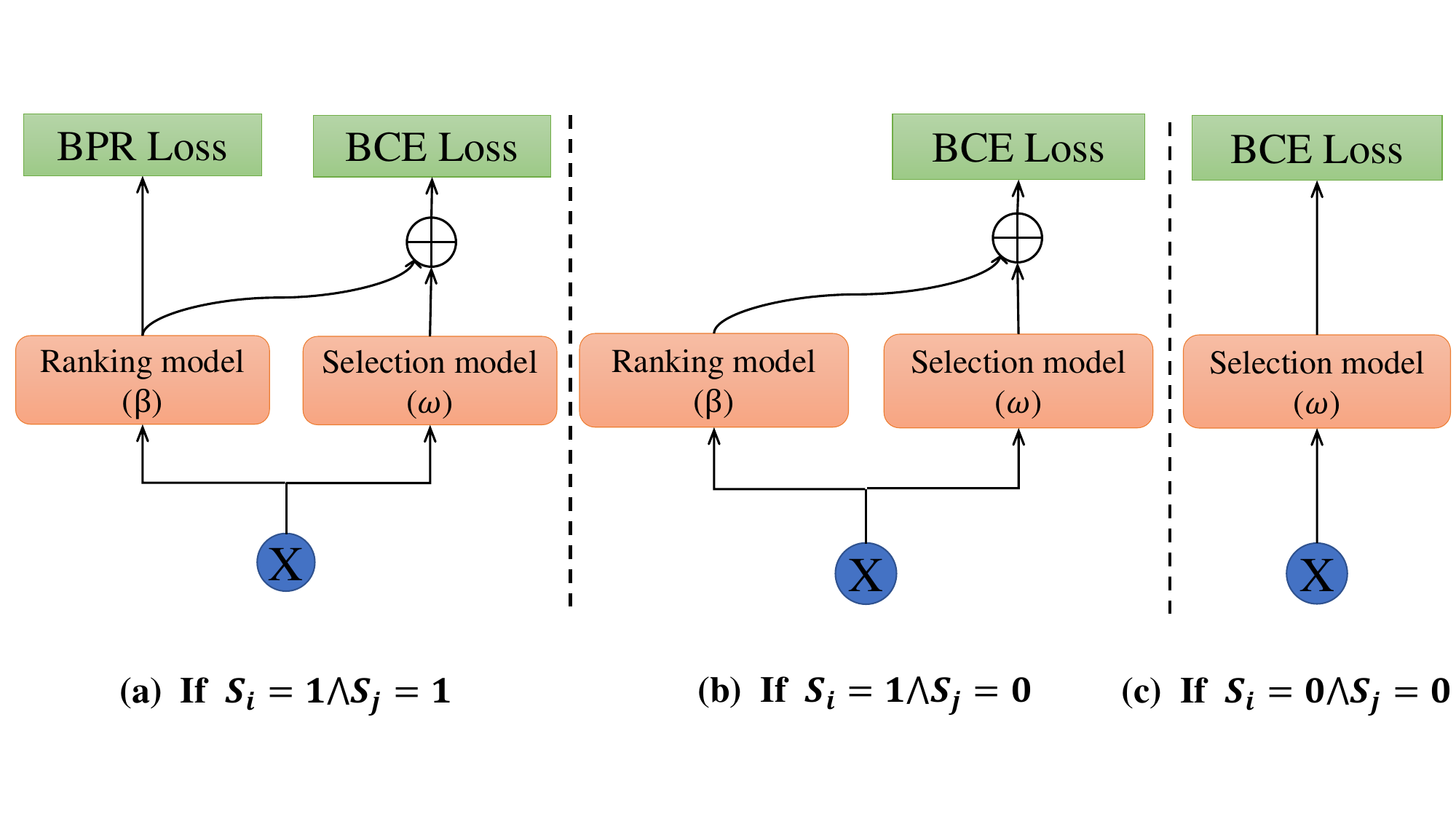}
    \caption{The unified debiasing model structure for optimizing pairwise neural format of CLD.}
    \label{fig:model}
\end{figure}

To conduct the optimization, we first parameterize the models with neural networks, as shown in Figure~\ref{fig:model}. Given a $(q,d)$ pair, its representation is denoted as $\mathbf{x}$.
Based on the representation, the relevance ranking model and selection model are defined as feed-forward neural networks, denoted as $f_{\bm{\beta}}(\cdot)$ and $f_{\bm{\omega}}(\cdot)$, respectively. 
Furthermore, we assume that in the second and third terms of Equation~\eqref{eq: nobothbias_likelihood_pair}, the selection of document pairs $(\mathbf{x}_i,\mathbf{x}_j)$ are independent:
\begin{align}\nonumber
        \mathrm{Pr}(s_{i},s_{j}|\bar{r}_{i}>\bar{r}_{j},\mathbf{x}_{i},\mathbf{x}_{j})
        &=\mathrm{Pr}(s_{i}|\bar{r}_{i}>\bar{r}_{j},\mathbf{x}_{i})\mathrm{Pr}(s_{j}|\bar{r}_{i}>\bar{r}_{j},\mathbf{x}_{j});
    \\\nonumber
        \mathrm{Pr}(s_{i},s_{j}|\mathbf{x}_{i},\mathbf{x}_{j})
        &= \mathrm{Pr}(s_{i}|\mathbf{x}_{i})\mathrm{Pr}(s_{j}|\mathbf{x}_{j}).
\end{align}

As shown in Figure~\ref{fig:model}(a), if both of the documents in a pair are selected into the top-$k$ positions, the likelihood of relevance part and conditional selection part can be formulated with BPR loss and Binary Cross Entropy loss, respectively. if only one document in a pair is selected, the conditional selection likelihood can be approximated as that shown in Figure~\ref{fig:model}(b). If neither of the two documents in a pair is selected, the selection likelihood can be formulated with Binary Cross Entropy loss directly (Figure~\ref{fig:model}(c)). 
Therefore, the overall pairwise objective function becomes:
\begin{equation}
    \label{eq: MULE_pair}
    \begin{aligned}
       &\hphantom{{}={}}\mathcal{O}_{\mathrm{CLD}}^{\mathrm{pair}}(\bm{\beta},\bm{\omega})=s_{i}s_{j}\log~\sigma\left(f_{\bm{\beta}}(\mathbf{x}_{i})-f_{\bm{\beta}}(\mathbf{x}_{j})\right)\\
       &+s_{i}\log~\sigma\left(f_{\bm{\omega}}(\mathbf{x}_{i})+f_{\bm{\beta}}(\mathbf{x}_{i})-f_{\bm{\beta}}(\mathbf{x}_{j})\right)+(1-s_{i})\log~\sigma\left(1-f_{\bm{\omega}}(\mathbf{x}_{i})\right)\\
       &+s_{j}\log~\sigma\left(f_{\bm{\omega}}(\mathbf{x}_{j})+f_{\bm{\beta}}(\mathbf{x}_{i})-f_{\bm{\beta}}(\mathbf{x}_{j})\right)+(1-s_{j})\log~\sigma\left(1-f_{\bm{\omega}}(\mathbf{x}_{j})\right),
    \end{aligned}
\end{equation}
where $\sigma$ denotes the sigmoid function. To maximize Equation~\eqref{eq: MULE_pair}, we can get the approximately unbiased estimation of $\bm{\beta}$ and $\bm{\omega}$:
\[
    (\bm{\beta^{*}},\bm{\omega^{*}}) \leftarrow \mathop{\arg\max}_{\bm{\beta},\bm{\omega}}~\mathcal{O}_{\mathrm{CLD}}^{\mathrm{pair}}(\bm{\beta},\bm{\omega}).
\]
Algorithm \ref{alg: pair_training} illustrates the optimization procedure for Equation~\eqref{eq: MULE_pair}.

As for online ranking, given a $(q,d)$ pair, its ranking score is calculated by the ranking model $f_{\bm{\beta^*}}$:
$$
\hat{r} = f_{\bm{\beta^*}}(\phi(q,d)).
$$


%% file: 5-experiment.tex
\section{Experiment setup}
We conducted experiments to evaluate the proposed CLD and its extension CLD$^\textrm{pair}$, by following the settings presented in the existing unbiased learning to rank studies~\citep{Joachims2017Unbiased,Ai2018Unbiased,Agarwal2019framework,Oosterhuis2020Policy}.

\textbf{Datasets}: Two widely used public datasets,  YaHooC14B~\citep{chapelle2011yahoo} and WEB10K~\citep{qin2013introducing}, were used in our experiments. YaHooC14B contains around 30,000 queries, each associated with averaged of 24 documents. Each query-document pair is depicted with a 700-dimension feature vector and five-grade relevance labels. 
WEB10K has 10,000 queries and each associated with about 125 documents. Each query-document pair is depicted with a 136-dimension feature vector and a five-grade relevance label. 
Following the practices in~\citep{Joachims2017Unbiased}, we converted the relevance label in both two datasets with $r=1$ for grades 3 and 4 and $r=0$ for the others.
Only the set 1 of YaHooC14B and the first fold of WEB10K was used for training. Expert annotated labels in the test sets were used to evaluate the ranking accuracy.

\textbf{Click simulation}: Following the practices in~\citep{Joachims2017Unbiased}, the users' interactions with search engines were simulated and got the clicks.  First, 1\% labeled data were randomly sampled from the dataset and used to train an SVM$^{rank}$\citep{Joachims2002Optimizing} as the production ranker. Then for each click session, a query was uniformly sampled and the ranking result was generated by the production ranker. To simulate users' click, 
the position based model (PBM) \citep{Richardson2007Predicting,Ai2021survey} was adopted in which $(E=1 \wedge R=1) \Rightarrow C=1$, a click occurs only when the document is examined and is relevant. For every $(q,d)$ pair, the examination probability is based on the displayed position:
\begin{equation}
\label{eq: examination_prob}    
   \mathrm{Pr}(E=1|P=k)=
   \begin{cases}
       \left(\frac{1}{k}\right)^{\eta}, & \mbox{if\;} k\leq K \\
       0, & \mathrm{else} 
   \end{cases}
\end{equation}
where $\eta$ is the parameter to control the severity of position bias, and $K$ is the cut-off position. The examination probability is also the propensity score in the proposed approach and we assume it is known in advance. During the process, the irrelevant documents were allowed to be clicked with a small probability to simulate the click noise.

\textbf{Baselines}: State-of-the-art unbiased learning to rank approaches were adopted as the baselines:
\begin{description}
\item[Naive]: Directly regarding the clicks as relevance labels.
\item[IPS~\citep{Joachims2017Unbiased}]: Correcting the position bias with propensity score.
\item[Heckman$^\textrm{rank}$~\citep{Ovaisi2020Correcting}]: Correcting the sample selection bias with Heckman two-stage method.
\item[RankAgg\citep{Ovaisi2020Correcting}]: Mitigating both the position bias and sample selection bias by combining the results of IPS and Heckman$^\textrm{rank}$.
\item[Oracle]: Using the non-discarded expert annotated labels to learn the ranking model. It showed the (theoretical) performance upper bound on the dataset.
\end{description}
Policy-aware IPS~\citep{Oosterhuis2020Policy}  was not chosen as a baseline because it assumes the previous ranking models should be stochastic, which violets the Assumption (1). 

\textbf{Evaluation metric}: NDCG@1, NDCG@3, and MAP were used to evaluate the accuracy of the baselines and the proposed method. 

\textbf{Implementation details}: Similar to existing studies~\citep{Ai2018Unbiased, Agarwal2019framework, Vardasbi2020When}, we used a three layers neural networks with $elu$ activation function as the ranking model for Naive, IPS, Oracle and CLD$^\textrm{pair}$, with the hidden sizes $[256,128,64]$, and dropout probability of $0.5$. For Heckman$^\textrm{rank}$ and CLD (pointwise and linear), the ranking model was set to linear. The selection models for CLD and CLD$^\textrm{pair}$ were also set to linear. The learning rate were tuned among $\{2e{-4},5e{-4},1e{-3},2e{-3},5e{-3}\}$. The $L2$ regularization was used and the trade-off factor was tuned between $[1e{-3}, 1e{-2}]$. The correlation $\gamma$ in Equation~\eqref{eq: MULE_point} was tuned between $[0.05, 0.30]$. In all of the experiments, the reported numbers were the averaged results after training 12 epochs with 5 different random seeds.

The source code, data, and experiments will be available at \url{https://github.com/hyz20/CLD.git}

%% file: 6-result.tex
\section{Results and discussions}
\label{sec: result and discussion}

Table~\ref{tab:result} shows the ranking accuracy of our approaches and the baselines, on YaHooC14B and WEB10K. The results showed that the proposed CLD and CLD$^\textrm{pair}$ outperformed the baselines in terms of NDCG and MAP. 
``Oracle'' is the upper bound of the performance, since it uses expert annotated labels. The results verified the effectiveness of the unified bias mitigation in top-k ranking. 

\begin{table}[t]
    \caption{Ranking accuracy on YaHooC14B and WEB10K. Boldface means the best performed approaches (excluding Oracle). Experimental settings: top-5 cut-off, $\mathbf{\eta=0.1}$, $10^5$ click sessions, and 10\% click noise. We also present the 90\% confidence interval of $t$-distribution for our methods.}
    \resizebox{0.5\textwidth}{!}{
        \begin{tabular}{@{}c@{  }|c@{ }c@{ }c|c@{ }c@{ }c@{}}
        \hline
            \multirow{2}*{Method} & \multicolumn{3}{c}{YaHooC14B} & \multicolumn{3}{c}{WEB10K}\\
            \cline{2-7}
            ~ & NDCG@1 & NDCG@3 & MAP & NDCG@1 & NDCG@3 & MAP \\
        \hline
            Naive & 0.606 & 0.593 & 0.592 & 0.383 & 0.350 & 0.294 \\
            IPS & 0.650 & 0.619 & 0.609 & 0.413 & 0.368 & 0.282	 \\
            $\mathrm{Heckman}^\mathrm{rank}$ & 0.608 & 0.590 & 0.587 & 0.350 & 0.331 & 0.287 \\
            RankAgg & 0.649 & 0.623 & 0.608 & 0.413 & 0.375 & 0.299 \\
        \hline
            $\mathrm{CLD}$ & $0.661\pm.002$ & $\mathbf{0.631\pm.001}$ & $\mathbf{0.616\pm.001}$ & $0.434\pm.005$ & $0.391\pm.003$ & $0.312\pm.001$ \\
            $\mathrm{CLD}^\mathrm{pair}$ & $\mathbf{0.662\pm.001}$ & $0.630\pm.001$ & $0.615\pm.001$ & $\mathbf{0.439\pm.001}$ & $\mathbf{0.397\pm.001}$ & $\mathbf{0.312\pm.000}$ \\
        \hline
            Oracle & 0.666 & 0.636 & 0.622 & 0.455 & 0.416 & 0.332 \\
        \hline
        \end{tabular}}
\label{tab:result}
\end{table}

\begin{figure*}[t]
    \includegraphics[width=\textwidth]{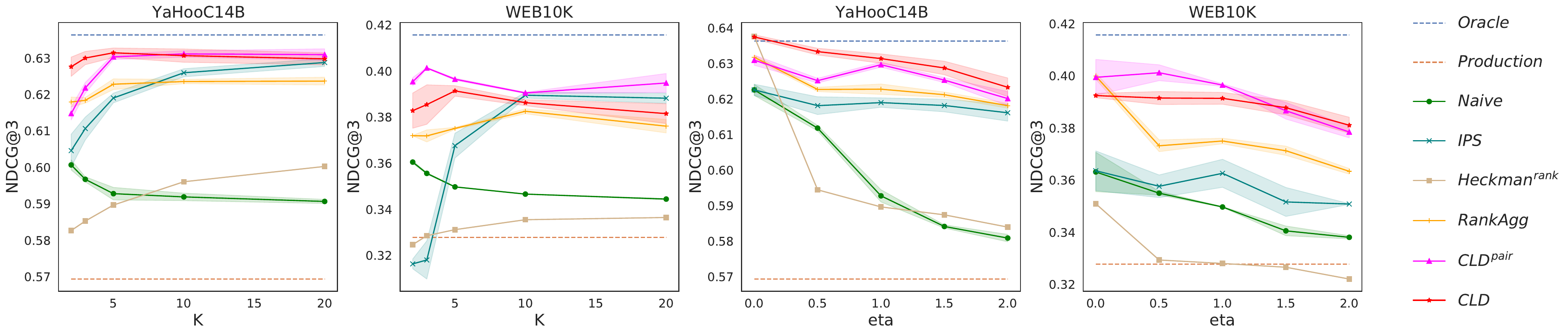}
    \caption{Performance curves of different methods w.r.t. bias severity levels. Experimental settings: $10^5$ click sessions, and 10\% click noise. Shaded area indicates the 90\% confidence intervals of $t$-distribution. \emph{Left two figures}: performance curves w.r.t. different severity of sampling selection bias. \emph{Right two figures}: performance curves w.r.t. different severity of position bias.}
    \label{fig:bias_result}
\end{figure*}

To further reveal how CLD and CLD$^\text{pair}$ outperformed the baselines, we conducted a group of exploratory experiments to answer the following research questions:
\begin{flushleft}
    \begin{itemize}
        \item[\textbf{RQ1}] How does CLD perform under different severity levels of sample selection bias and position bias?
        \item[\textbf{RQ2}] How does CLD perform under different scales of click data?
        \item[\textbf{RQ3}] Is CLD robust to click noise?
        \item[\textbf{RQ4}] Is CLD robust to misspecified propensity score?
        \item[\textbf{RQ5}] How does CLD perform with different base model?
    \end{itemize}
\end{flushleft}



\subsection{The effect of biases severity (RQ1)}
To varying the severity levels of sample selection bias, we changed the ranking cut-off position $k$ from 2 to 20. Smaller $k$ leads to more severe sample selection bias. The left two sub-figures of Figure~(\ref{fig:bias_result}) show the performance curves 
of different approaches w.r.t. different $k$ values.  
From the results, we can see that in general CLD and CLD$^\textrm{pair}$ outperformed the baselines at all of the $k$ values (except CLD when $k>10$ on WEB10K). On both datasets, when $k$ was small 
, CLD and CLD$^\textrm{pair}$ outperformed the baselines with a large margin and achieved the performance closing to the upper bound. With the increasing of $k$, the improvements of CLD and CLD$^\textrm{pair}$ over IPS gradually become limited. This is because sample selection bias gets milder for larger $k$, making position bias dominates the negative effects of bias. Similar performance curves also came to RankAgg, another model which can mitigate both position bias and sample selection bias. 

On the contrary, increasing $k$ will lead to the performance drop of naive methods. This is because the naive method can handle neither of these two biases. Increasing data will not further improve its performance but decrease its performance instead. Also note that CLD$^\textrm{pair}$ outperformed CLD on WEB10K since it learns a neural ranking model based on the pairwise loss. However, all methods can achieve relatively high performances on YaHooC14B. The spaces for further improvement are limited, leading to similar performances for CLD$^\textrm{pair}$ and CLD.


To change the severity levels of position bias, we tuned the the parameter $\eta$ in Equation~\eqref{eq: examination_prob} from 0.0 to 2.0. Larger $\eta$ leads to more severe position bias. The right two sub-figures in Figure~\ref{fig:bias_result} illustrate the performances curves 
w.r.t. different $\eta$ values.
From the results, we can see that CLD and CLD$^\textrm{pair}$ still outperformed all the baselines on both datasets.
With the increasing of $\eta$, the methods that can correct position bias (except Heckman$^\textrm{rank}$ and Naive) have slight performance drops. Among the baselines, Heckman$^\textrm{rank}$ achieved the higher performance when $\eta=0$ (no position bias), but dropped rapidly when $\eta$ increases. RankAgg also suffered from the performance drop with the increasing $\eta$ because it is an ensemble of Heckman$^\textrm{rank}$. The phenomenon confirmed the conclusion in Section~\ref{sec: causalview_position}: only mitigating one bias separately still leads to a biased result in top-$k$ ranking. Simply aggregating the results outputted by the methods that only correct one bias still leads to biased results.

\begin{figure}[t]
    \includegraphics[scale=0.21]{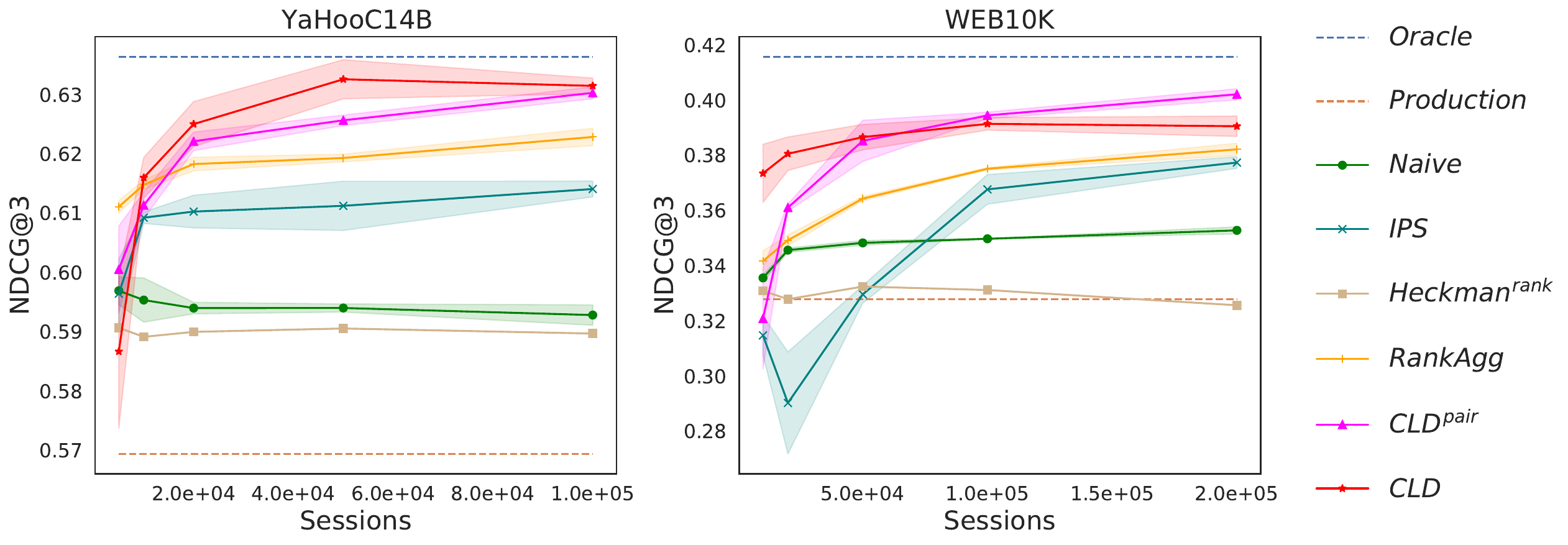}
    \caption{Performance curves of different methods w.r.t. the number of click sessions. Experimental settings: top-5 cut-off, $\mathbf{\eta=1.0}$ and 10\% click noise. }
    \label{fig:scale_result}
\end{figure}

\begin{figure}[t]
    \includegraphics[scale=0.21]{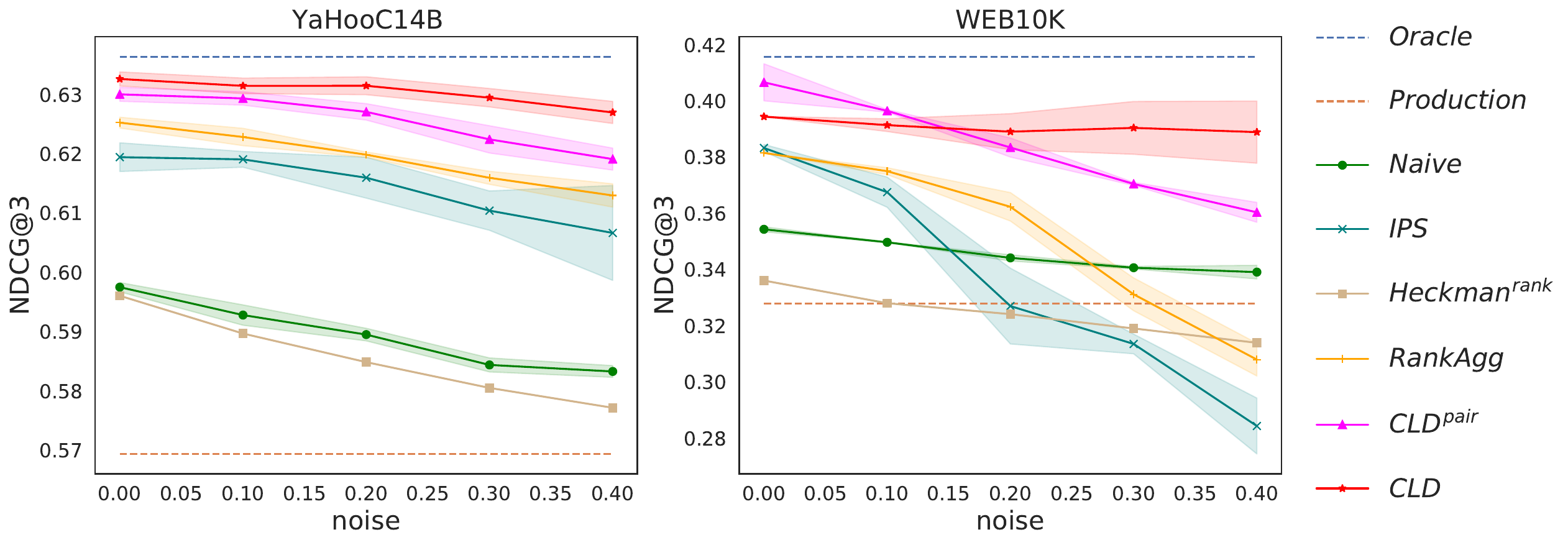}
    \caption{Performance curves of different methods w.r.t. click noise severity levels. Experimental settings: top-5 cut-off, $\mathbf{\eta=1.0}$ and trained with $10^5$ click sessions. }
    \label{fig:noise_result}
\end{figure}

\subsection{The effects of click scales (RQ2)}
We tested the performances of different methods by varying the scale of the click data. Figure~\ref{fig:scale_result} illustrates the performance curves of different methods w.r.t. the number of click sessions used for training the models. The results indicate that both CLD and CLD$^\textrm{pair}$ consistently outperformed the baseline methods over different click session scales. According to Equation~\eqref{eq: MULE_point} and \eqref{eq: MULE_pair}, CLD and CLD$^\textrm{pair}$ have the ability of utilizing the unobserved data in training. The ability makes them perform well even being trained with a limited number of click sessions. With more click sessions being involved in training, the performances of CLD and CLD$^\textrm{pair}$ steadily improved. 
In contrast, IPS and Heckman$^\textrm{rank}$ can only correct one bias in top-$k$ ranking. Therefore, with the increasing number of click sessions used for training, they underperformed those methods that can mitigate both position bias and sample selection bias (e.g., RankAgg, CLD and CLD$^\textrm{pair}$). All these results clearly verified the advantages of the proposed unified bias mitigation. 

\subsection{The effects of click noises (RQ3)}
We also conducted experiments with variant noise levels in the clicks, by changing the probability of clicking irrelevant documents from 0.0 to 0.5 when generating the training data. According to the results shown in Figure~\ref{fig:noise_result}, both CLD and CLD$^\textrm{pair}$ outperformed the baselines at different noise levels, indicating the robustness of the unified bias mitigation approach. Particularly, among all of the methods, CLD has minimal performance drops. 

We analyzed the reasons and found that each noise click will produce more mistake pairs in pairwise methods than that of in pointwise methods. Therefore, when training with these mistake pairs, pairwise method will be suffered more. However, for pointwise method, each noise click is only presented once in the training set, making it more robust than the pairwise models. 

\begin{figure}[t]
    \includegraphics[scale=0.21]{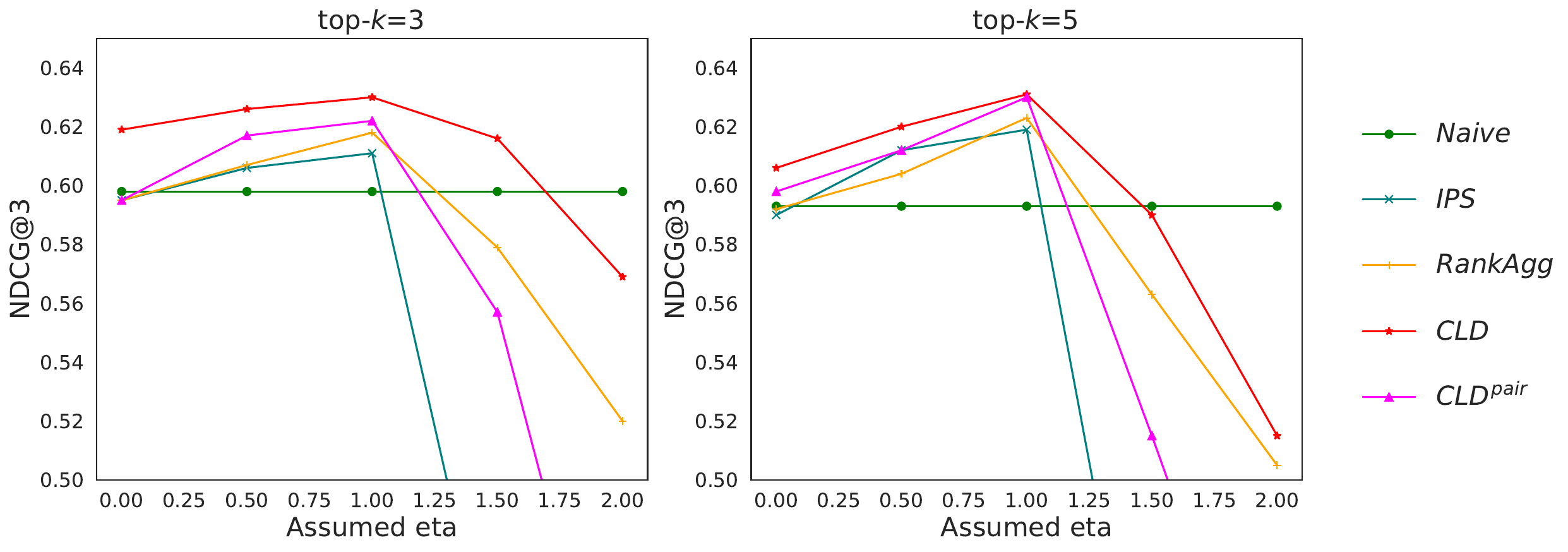}
    \caption{Performance curves of different methods on YaHooC14B w.r.t. degrees of misspecified propensity scores in different top-k cut-offs. The true $\mathbf{\eta}=1$ and click noise is 10\%. }
    \label{fig: mis ips}
\end{figure}

\subsection{\mbox{Effects of misspecified propensity score (RQ4)}}
The result we reported before assumes that the model knows the true propensity score, which is often difficult in the real world. 
In this experiment, we conducted experiments to test the performance of each method under various degrees on misspecified propensity scores and different top-$k$ cut-offs, characterized by parameters $\eta$ and $k$, respectively. The true value $\eta = 1$ and we varied it in $[0.0, 2.0]$. We tested the cases when $k=3$ and $k=5$. Note that Heckman$^\textrm{rank}$ are not considered as a baseline in this experiment. This is because Heckman$^\textrm{rank}$ does not use propensity scores.

Figure~\ref{fig: mis ips} illustrates the performance curves of CLD, CLD$^{pair}$, IPS, and RankAgg on YaHooC14B, under various degrees of misspecified propensity scores. The left and right figures respectively illustrate the results when $k=3$ and $k=5$. From the results, we can see that in general CLD outperformed the best in all degrees of misspecified propensity scores, which indicates the robustness of CLD. When the propensity was overestimated (i.e., $\eta<1$), all methods related to propensity score only have a slight performance drop. However, all methods have violent performance drops if the propensity was underestimated (i.e., $\eta>1$). This is because when the propensity is underestimated, the estimated propensity becomes smaller than its true value, and thus increasing the variance of propensity re-weighting. Even when the propensity was underestimated, the proposed CLD still outperformed other methods with large margins.This is attributed to CLD avoids dividing by propensity score in the whole loss function and therefore can reduce the variance caused by the underestimation of the propensity. Moreover, we found that the effects of misspecified propensity scores were more severe on large $k$. This is because the larger the ranking positions, the more suffers come from the position bias.


\begin{table}[t]
    \caption{Ranking accuracy comparison among different variants of CLD on YaHooC14B and WEB10K. Boldface means the best performed approaches (excluding Oracle). Experimental settings: top-5 cut-off, $\mathbf{\eta=0.1}$, $10^5$ click sessions, and 10\% click noise.}
    \label{tab:compare}
    \resizebox{0.45\textwidth}{!}{
        \begin{tabular}{@{}c@{  }|c@{ }c@{ }c|c@{ }c@{ }c@{}}
        \hline
            \multirow{2}*{Method} & \multicolumn{3}{c}{YaHooC14B} & \multicolumn{3}{c}{WEB10K}\\
            \cline{2-7}
            ~ & NDCG@1 & NDCG@3 & MAP & NDCG@1 & NDCG@3 & MAP \\
        \hline
            CLD & 0.661 & 0.631 & 0.616 & 0.434 & 0.391 & \textbf{0.312} \\
            CLD-N & 0.652 & 0.619 & 0.610 & 0.340 & 0.321 & 0.284	 \\
        \hline
            CLD$^\textrm{pair}$ & \textbf{0.662} & 0.630 & 0.615 & \textbf{0.439} & \textbf{0.397} & \textbf{0.312} \\
            CLD$^\textrm{pair}$-L & 0.660 & \textbf{0.634} & \textbf{0.618} & 0.431 & 0.389 & 0.309 \\
        \hline
            Oracle & 0.666 & 0.636 & 0.622 & 0.455 & 0.416 & 0.332 \\
        \hline
        \end{tabular}}
\end{table}

\subsection{The effects of base model (RQ5)}
In previous experiments, CLD was designed as a linear ranking model because of its theoretic grantees, while CLD$^\textrm{pair}$ was designed to use nonlinear neural networks as its ranker. In this experiment, we modified these models so that CLD was based on a neural network with three hidden layers and CLD$^\textrm{pair}$ used a linear model as the ranker, denoted as CLD-N and CLD$^\textrm{pair}$-L, respectively. 

Table~\ref{tab:compare} reports the ranking accuracy of CLD, CLD$^\textrm{pair}$, and their variations, on YaHooC14B and WEB10K. From the results, we found that (1) CLD-N performed worst among these methods, especially on WEB10K. Compared to CLD, CLD-N used a nonlinear neural network as its ranker, which makes it lose the theoretical guarantees; (2) CLD$^\textrm{pair}$ outperformed CLD$^\textrm{pair}$-L on WEB10K and performed comparably on YaHooC14B. Please note that WEB10K is larger than YaHooC14B and all methods can achieve relatively scores on YaHooC14B. We concluded that using a linear model in CLD and using nonlinear neural networks in CLD$^\textrm{pair}$ are reasonable settings. 


%% file: 7-Appendix.tex
\section{Appendix}
\subsection{Derivation of the lower bound of CLD}
\label{appendix:derivation cld lb}
In section~\ref{sec: Optimization}, we have derived the parameterized log-likelihood CLD:
\begin{equation}
    \begin{split}
        \mathcal{L}_{\mathrm{CLD}}(\bm{\beta},\bm{\omega}) =& -\sum_{i=1\land s_{i}=1}^{N_s} \left(\mathbb{E}\left[\frac{c_i}{\rho_i}\right]-\mathbf{x}_{i}^{T}\bm{\beta} \right)^2\\
        &+\sum_{i=1\land s_{i}=1}^{N_s}\log\Phi\left(\frac{\mathbf{x}_{i}^{T}\bm{\omega}+\gamma \left(\mathbb{E}\left[\frac{c_i}{\rho_i}\right]-\mathbf{x}_{i}^{T}\bm{\beta} \right)}{(1-\gamma^{2})^{\frac{1}{2}}}\right)\\
        &+\sum_{i=1\land s_{i}=0}^{N_u}\log\left(1-\Phi(\mathbf{x}_{i}^{T}\bm{\omega})\right),
    \end{split}
    \label{eq: MULE_point}
\end{equation}
However, this equation contains the expectation of propensity reweighted click $\mathbb{E}\left[\frac{c_i}{\rho_i}\right]$, which need to be pre-calculated.

In our implementation, the expectation $\mathbb{E}\left[\frac{c_i}{\rho_i}\right]$ can be substituted by the propensity reweighted click $\frac{c_i}{\rho_i}$. Then the Equation~\eqref{eq: MULE_point} can be transformed as:

\begin{equation}
    \begin{split}
        \mathcal{L}_{\mathrm{CLD}}^{'}(\bm{\beta},\bm{\omega}) =& -\sum_{i=1\land s_{i}=1}^{N_s} \left(\frac{c_i}{\rho_i}-\mathbf{x}_{i}^{T}\bm{\beta} \right)^2\\
        &+\sum_{i=1\land s_{i}=1}^{N_s}\log\Phi\left(\frac{\mathbf{x}_{i}^{T}\bm{\omega}+\gamma \left(\frac{c_i}{\rho_i}-\mathbf{x}_{i}^{T}\bm{\beta} \right)}{(1-\gamma^{2})^{\frac{1}{2}}}\right)\\
        &+\sum_{i=1\land s_{i}=0}^{N_u}\log\left(1-\Phi(\mathbf{x}_{i}^{T}\bm{\omega})\right),
    \end{split}
    \label{eq: MULE_point_LB}
\end{equation}

Here we will show the expectation of this equation is lower bound of Equation~\eqref{eq: MULE_point}. First, we have the following lemmas:
\begin{lemma}
For a function $f(x) = h(x) + g(x)$, $f$ is convex if both $h$ and $g$ are convex.
\label{lm: add}
\end{lemma}

\begin{lemma}
For a composite function $f(x) = h(g(x))$, $f$ is convex if $h$ is convex and nondecreasing, and $g$ is convex.
\label{lm: composite}
\end{lemma}
In Equation~\ref{eq: MULE_point_LB}, we can see that the first term is obviously a convex function on $\frac{c_i}{\rho_i}$. As for the second term, the $\log$ function is a convex and nondecreasing function, the cumulative distribution function $\Phi$ is also a convex function when random variable takes a value greater than 0. Based on above lemma~\ref{lm: composite}, the second term is also a convex function on $\frac{c_i}{\rho_i}$. Since the first term and the second term are both the convex functions on $\frac{c_i}{\rho_i}$, Equation~\eqref{eq: MULE_point_LB} is a convex function on $\frac{c_i}{\rho_i}$ based on lemma~\ref{lm: add}.

With the help of Jensen’s inequality, the expectation of Equation~\eqref{eq: MULE_point_LB} on $\frac{c_i}{\rho_i}$ can be proved to be the lower bound of Equation~\eqref{eq: MULE_point}:
\begin{equation}
    \begin{split}
    \mathbb{E}_{\frac{c_i}{\rho_i}}\left[\mathcal{L}_{\mathrm{CLD}}^{'}(\bm{\beta},\bm{\omega})\right]\leq\mathcal{L}_{\mathrm{CLD}}(\bm{\beta},\bm{\omega})
    \end{split}
\end{equation}
Since our CLD is to maximize the whole likelihood of observed data, we can maximize the lower bound of this likelihood as illustrated in this equation.

\subsection{Proof of Theorem 2}
\label{appendix:proof of theorem 2}
\begin{proof}
The likelihood defined in Equation~\eqref{eq: MULE_point} 
can be rewritten as the form for a single sample:
\[
    \begin{split}
        \mathcal{L}_{\mathrm{CLD}} &= \sum_i^{\mathcal{D}} s_i\left( \log\Phi\left(\frac{\mathbf{x}_{i}^{T}\bm{\omega}+\gamma \epsilon}{(1-\gamma^{2})^{\frac{1}{2}}}\right)-\epsilon^2\right)+(1-s_i)\log\left(1-\Phi(\mathbf{x}_{i}^{T}\bm{\omega})\right),
    \end{split}
    \label{eq: single_loss}
\]
where $\epsilon= \left(\mathbb{E}\left[\frac{c_i}{\rho_i}\right]-\mathbf{x}_{i}^{T}\bm{\beta}\right)$, for the ease of notation. Since each record in dataset $\mathcal{D}$ is independent, we have:
\[
    \begin{split}
        \mathbb{V}\left[\mathcal{L}_{\mathrm{CLD}}\right] &= \mathbb{V}\left[\sum_i^{\mathcal{D}} s_i\left( \log\Phi\left(\frac{\mathbf{x}_{i}^{T}\bm{\omega}+\gamma \epsilon}{(1-\gamma^{2})^{\frac{1}{2}}}\right)-\epsilon^2-\log\left(1-\Phi(\mathbf{x}_{i}^{T}\bm{\omega})\right)\right)\right]\\
        &=\sum_i^{\mathcal{D}} \mathbb{V}\left[s_i\left( \log\Phi\left(\frac{\mathbf{x}_{i}^{T}\bm{\omega}+\gamma \epsilon}{(1-\gamma^{2})^{\frac{1}{2}}}\right)-\epsilon^2-\log\left(1-\Phi(\mathbf{x}_{i}^{T}\bm{\omega})\right)\right)\right]\\
        &=\sum_i^{\mathcal{D}} \mathbb{V}\left[s_i\right]\left( \log\Phi\left(\frac{\mathbf{x}_{i}^{T}\bm{\omega}+\gamma \epsilon}{(1-\gamma^{2})^{\frac{1}{2}}}\right)-\epsilon^2-\log\left(1-\Phi(\mathbf{x}_{i}^{T}\bm{\omega})\right)\right)^2
    \end{split}
    \label{eq: variance_derivation}
\]
\end{proof}
It is worth noting that $s_i$ is a random variable that obeys the Bernoulli distribution. The other part except $s_i$ can be treated as a constant for a given $i$, which is related to input $\mathbf{x_i}$. Therefore, the variance of CLD depends on the variance of $s_i$ and the scale of input $\mathbf{x_i}$. In comparison with IPS, the variance of CLD avoids dividing by propensity, thus avoiding being affected by those extreme minimal propensity values.

\subsection{Derivation of $\mathrm{CLD^{pair}}$}
\label{appendix:derivation cld pair}
To derive the pairwise format of CLD, we first give the unbiased log-likelihood in pairwise format:
\begin{equation}
    \begin{split}
        \mathcal{L}_{\mathrm{unbiased}}^{\mathrm{pair}} &= \sum_{r_{i}>r_{j}}\log\left(\mathrm{Pr}(r_{i}>r_{j}|\mathbf{x}_{i},\mathbf{x}_{j})\right).
    \end{split}
    \label{eq: unbias_likelihood_pair}
\end{equation}
For a document $i$ and document $j$ in the ranking list of query $\bm{q}$, the pairwise unbiased likelihood is consists of the relative order of their relevance comparisons. Unfortunately, the relevance $r_{i}$ and $r_{j}$ is unknown for us. What we can observe is the click signal of each document, and the pairwise log-likelihood of these observed data can be written as:
\begin{equation}
    \begin{split}
        \mathcal{L}_{\mathrm{naive}}^{\mathrm{pair}} &= \sum_{c_{i}>c_{j},\;s_{i}=1\land s_{j}=1}\log\left(\mathrm{Pr}(c_{i}>c_{j}, s_{i}, s_{j}|\mathbf{x}_{i},\mathbf{x}_{j})\right)\\
        &+\sum_{s_{i}=0\lor s_{j}=0}\log\left(\mathrm{Pr}(s_{i},s_{j}|\mathbf{x}_{i},\mathbf{x}_{j})\right).
    \end{split}
    \label{eq: naive_likelihood_pair}
\end{equation}
Though it can be optimized, the naive log-likelihood suffers from both position bias from $c_{i}$ and the sample selection bias from $s_{i}$. There is an obvious gap between the unbiased log-likelihood and the naive log-likelihood. To this end, we first detach the position bias from click signal with the help propensity score:
\begin{equation}
    \begin{split}
        \mathcal{L}_{\mathrm{de. posi.}}^{\mathrm{pair}} &= \sum_{\bar{r}_{i}>\bar{r}_{j},\;s_{i}=1\land s_{j}=1}\log\left(\mathrm{Pr}(\bar{r}_{i}>\bar{r}_{j}, s_{i}, s_{j}|\mathbf{x}_{i},\mathbf{x}_{j})\right)\\
        &+\sum_{s_{i}=0\lor s_{j}=0}\log\left(\mathrm{Pr}(s_{i},s_{j}|\mathbf{x}_{i},\mathbf{x}_{j})\right).
    \end{split}
    \label{eq: noposi_likelihood_pair}
\end{equation}
where $\bar{r}_i=\mathbb{E}\left[\frac{c_i}{\rho_i}\right]$ and $\bar{r}_j=\mathbb{E}\left[\frac{c_j}{\rho_j}\right]$ for simplifying the notations, Equation~\eqref{eq: noposi_likelihood_pair} still suffers sample selection bias because the first term still contains $s_i$. As have shown in Section~3.2, the likelihood of selected data can be decomposed to contain the unbiased learning target, thus detaching sample selection bias: 
\begin{align}
        \mathcal{L}_{\mathrm{de.}\;\mathrm{both}\;\mathrm{biases}}^{\mathrm{pair}} &= \sum_{\bar{r}_{i}>\bar{r}_{j},\;s_{i}=1\land s_{j}=1}\log\left(\mathrm{Pr}(\bar{r}_{i}>\bar{r}_{j}|\mathbf{x}_{i},\mathbf{x}_{j})\right) \nonumber\\
        &+\sum_{\bar{r}_{i}>\bar{r}_{j},\;s_{i}=1\land s_{j}=1}\log\left(\mathrm{Pr}(s_{i},s_{j}|\bar{r}_{i}>\bar{r}_{j},\mathbf{x}_{i},\mathbf{x}_{j})\right) \nonumber\\
        &+\sum_{s_{i}=0\lor s_{j}=0}\log\left(\mathrm{Pr}(s_{i},s_{j}|\mathbf{x}_{i},\mathbf{x}_{j})\right), 
\end{align}
Now we have detached both position bias and sample selection bias from the pairwise naive log-likelihood. To maximize the above log-likelihood, we can obtain an unbiased ranking model via the first term.

\subsection{More Experiment Results}
\label{appendix: add experiment}
We conducted more experiments to verify the robustness of CLD by misspecifying the propensity scores and changing the base model of CLD.   
\begin{figure}[t]
    \includegraphics[scale=0.21]{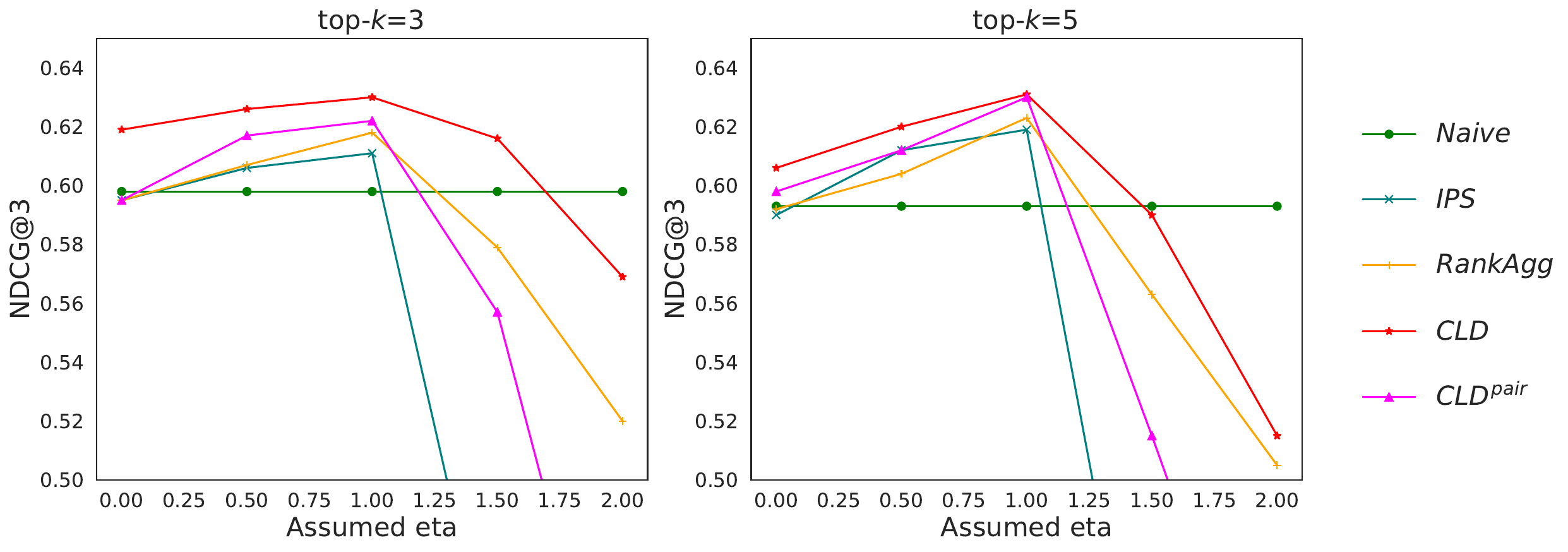}
    \caption{Performance curves of different methods on YaHooC14B w.r.t. degrees of misspecified propensity scores in different top-k cut-offs. The true $\mathbf{\eta}=1$ and click noise is 10\%. }
    \label{fig: mis ips}
\end{figure}
\subsubsection{The effect of misspecified propensity scores}
The result we reported in Section~\ref{sec: result and discussion} assumes that the model knows the true propensity score, which is often difficult in the real world. 
In this experiment, we conducted experiments to test the performance of each method under various degrees on misspecified propensity scores and different top-$k$ cut-offs, characterized by parameters $\eta$ and $k$, respectively. The true value $\eta = 1$ and we varied it in $[0.0, 2.0]$. We tested the cases when $k=3$ and $k=5$. Note that Heckman$^\textrm{rank}$ are not considered as a baseline in this experiment. This is because Heckman$^\textrm{rank}$ does not use propensity scores.

Figure~\ref{fig: mis ips} illustrates the performance curves of CLD, CLD$^{pair}$, IPS, and RankAgg on YaHooC14B, under various degrees of misspecified propensity scores. The left and right figures respectively illustrate the results when $k=3$ and $k=5$. From the results, we can see that in general CLD outperformed the best in all degrees of misspecified propensity scores, which indicates the robustness of CLD. When the propensity was overestimated (i.e., $\eta<1$), all methods related to propensity score only have a slight performance drop. However, all methods have violent performance drops if the propensity was underestimated (i.e., $\eta>1$). This is because when the propensity is underestimated, the estimated propensity becomes smaller than its true value, and thus increasing the variance of propensity re-weighting. Even when the propensity was underestimated, the proposed CLD still outperformed other methods with large margins. As have stated in Theorem~\ref{thm:variance}, the variance of CLD avoids dividing by propensity score and therefore can reduce the variance caused by the underestimation of the propensity. Moreover, we found that the effects of misspecified propensity scores were more severe on large $k$. This is because the larger the ranking positions, the more suffers come from the position bias.


\begin{table}[t]
    \caption{Ranking accuracy comparison among different variants of CLD on YaHooC14B and WEB10K. Boldface means the best performed approaches (excluding Oracle). Experimental settings: top-5 cut-off, $\mathbf{\eta=0.1}$, $10^5$ click sessions, and 10\% click noise.}
    \label{tab:compare}
    \resizebox{0.5\textwidth}{!}{
        \begin{tabular}{@{}c@{  }|c@{ }c@{ }c|c@{ }c@{ }c@{}}
        \hline
            \multirow{2}*{Method} & \multicolumn{3}{c}{YaHooC14B} & \multicolumn{3}{c}{WEB10K}\\
            \cline{2-7}
            ~ & NDCG@1 & NDCG@3 & MAP & NDCG@1 & NDCG@3 & MAP \\
        \hline
            CLD & 0.661 & 0.631 & 0.616 & 0.434 & 0.391 & \textbf{0.312} \\
            CLD-N & 0.652 & 0.619 & 0.610 & 0.340 & 0.321 & 0.284	 \\
        \hline
            CLD$^\textrm{pair}$ & \textbf{0.662} & 0.630 & 0.615 & \textbf{0.439} & \textbf{0.397} & \textbf{0.312} \\
            CLD$^\textrm{pair}$-L & 0.660 & \textbf{0.634} & \textbf{0.618} & 0.431 & 0.389 & 0.309 \\
        \hline
            Oracle & 0.666 & 0.636 & 0.622 & 0.455 & 0.416 & 0.332 \\
        \hline
        \end{tabular}}
\end{table}

\subsubsection{The effects of base model}
In previous experiments, CLD was designed as a linear ranking model because of its theoretic grantees, while CLD$^\textrm{pair}$ was designed to use nonlinear neural networks as its ranker. In this experiment, we modified these models so that CLD was based on a neural network with three hidden layers and CLD$^\textrm{pair}$ used a linear model as the ranker, denoted as CLD-N and CLD$^\textrm{pair}$-L, respectively. 

Table~\ref{tab:compare} reports the ranking accuracy of CLD, CLD$^\textrm{pair}$, and their variations, on YaHooC14B and WEB10K. From the results, we found that (1) CLD-N performed worst among these methods, especially on WEB10K. Compared to CLD, CLD-N used a nonlinear neural network as its ranker, which makes it lose the theoretical guarantees; (2) CLD$^\textrm{pair}$ outperformed CLD$^\textrm{pair}$-L on WEB10K and performed comparably on YaHooC14B. Please note that WEB10K is much larger than YaHooC14B and all methods can achieve relatively high scores on YaHooC14B. We concluded that using a linear model in CLD and using nonlinear neural networks in CLD$^\textrm{pair}$ are reasonable settings. 
